\documentclass[prl,twocolumn,showpacs]{revtex4}
\usepackage{graphicx}%

\begin{document}

\title{Finite Coherence Length of Thermal Noise in Percolating Systems}
\author{K. K. Bardhan~}
\author{C. D. Mukherjee}

\affiliation{Saha Institute of Nuclear Physics, 1/AF Bidhannagar, Calcutta
700 064, India}

\date{\today}

\begin{abstract}
Noise has been measured in two types of coductor-insulator mixtures as a
function of bias and composition. It was marked by a huge increase in
magnitude as the resistance increased only slightly due to Joule heating.
The noise (resistance) current scale $I_s$($I_r$) for nonlinearity were
found to scale with the linear resistance $R_o$ as $I_s(I_r) \sim {{R_o}^
{-x_s(x_r) }}$ where the exponent $x_s$ is equal to 0.80 and 0.68 in
carbon-wax and carbon-polyethylene respectively and $x_r \approx 0.5$. It is
shown that the large increase of noise in nonohmic regime as well as the
differences between the noise and resistance exponents are due to the
finite-sized inequilibrium thermal fluctuations whose coherence length is
same as the correlation length of the underlying percolating systems. A
expression for $x_s$ is derived.
\end{abstract}

\pacs{72.70.+m, 72.20.Ht, 72.60.+g}

\maketitle

The low frequency resistance noise, also known as $1/f$-noise, is a very
common phenomenon\cite {kogan} in normal conductors and is increasingly
being used as a tool for studying, particularly, disordered systems\cite
{weiss2}. While its origin remains controversial, some of its features are
well-established. In particular, the correlation length of resistance
fluctuations ($\xi_s$) is commonly assumed to be of the order of microscopic
lengths\cite {kogan}. This implies that the noise power should be inversely
proportional to the number of fluctuators or the system volume $L^d$, $d$
being the system dimensionality. But there are cases where
relatively large increases in the noise levels under certain experimental
conditions have been explained in terms of coarsening of the coherent volume
with correlation lengths increasing to macroscopic scale. This, other things
remaining same, amounts to an increase in the noise level by a factor of
${[L/(L/ \xi_s)]}^d = \xi_s^d$ with $\xi_s \le L$. For example, the large
increase in the noise at the onset of charge density wave motion in certain
conductors upon application of an electric field has been ascribed to the
finite coherence length of a charge density wave domain\cite {sabya}.
Analogous phenomena are the well-know critical opalescence in various
systems\cite {munster} where the static correlation length diverges as a
critical point is approached. Interestingly, in systems in nonequilibrium
states but not necessarily near any critical point, the correlation length
of fluctuations at equal time is also of macroscopic order\cite {schmitz}.

Earlier\cite {mbh}, the resistance behaviour of various composite samples
(random mixtures of conductors and insulators) exhibiting substantial Joule
heating was invesigated as a function of biasing currents. At low currents,
the resistance $R=V/I$ had a bias-independent value $R_o$ but increased at
high currents. It was found that the current $I_r$ at which the sample
resistance starts increasing scaled with $R_o$ as $I_r \sim R_o^{-x_r}$
where $x_r \approx 0.5$ is the onset exponent for resistance. In this
letter, we report on systematic measurements of low frequency resistance
noise as a function of currents in two composites systems of carbon-wax
(C-W) and carbon-polymer (C-HDPE). Samples under fixed currents were in
nonequilibrium steady states (below breakdown) with temperature gradients.
It was found that (i) the relative noise ${\cal S}$ (see below) behaved in a
similar fashion as the resistance such that the current scale $I_s$ for
onset of nonlinearity in noise\cite {rem3} and the associated onset exponent
$x_s$ could be defined as before: $I_s \sim R_o^{-x_s}$. $x_s$ was found to
be larger than $x_r$: \begin{eqnarray} \label{eq:xs_exp}
x_s = \left\{ \begin{array}{ll}
	0.80 \pm 0.03~{\rm (C\!-\!W)} \\
	0.68 \pm 0.04~{\rm (C\!-\!HDPE)}
              \end{array}
      \right.
\end{eqnarray}
(ii) The noise ${\cal S}$ of a sample as a function of resistance increased
tremendously compared to that caused by varying static disorder (i.e.
composition). It is argued that such increase in noise reslults from
additional resistance fluctuations induced by thermal fluctuations which are
coherent over a length of the correlation length $\xi$ of the underlying
percolation structure i.e. $\xi_s \approx \xi$. It is shown that this
assumption leads to an expression for $x_s$:
\begin{equation}
\label{eq:xs_the} x_s = x_r + d \nu /4t - w_J/4
\end{equation}
where $\nu$ and $t$ are percolation correlation and conductivity exponents
respectively\cite {stauffer} and \(w_{J} \) is the noise exponent\cite
{mbh}. Using $d=3$, $\nu = 0.9$\cite {stauffer} and experimental values
of $x_r$, $t$ and $w_J$ in (\ref {eq:xs_the}) yields $x_s = 0.83 \pm 0.10$
and $0.67 \pm 0.05$ for C-W and C-HDPE respectively. These values agree well
with the experimental ones in (\ref {eq:xs_exp}).

Noise have been studied extensively in many percolating systems like
composites both experimentally and theoretically but mostly in ohmic
regimes\cite {tremblay}. Recently, the bias-dependent behaviour of noise
near the percolation threshold has been measured \cite {nmb}. A composite
sample is primarily characterised by its conductor fraction $p$. Above a
particular value $p_c$, called percolation threshold, a continuous path is
formed from one end of the sample to another, enabling current to flow
through the system. A sample possesses a natural length scale which is the
two-particle correlation length $\xi$ given by $\xi \sim {(p-p_c)}^
{-\nu}$. The noise (in the ohmic regime) tends to diverge\cite {cc} as the
conducting fraction $(p)$ is reduced from large values above $p_c$. However,
this divergence has been explained to originate entirely from the static
geometrical disorder\cite {rammal}. $\xi_s$ remains of the order of
microscopic scales and independent of $\xi(p)$. Electrical conductivity of a
composite sample is characterised by the exponent $t$, defined by $R_o \sim
{(p-p_c)}^{-t}$\cite {stauffer}. The resistance $R_o$ decreases with
increasing $p$. The noise exponent $w_J$ is defined by \( {\cal S}(I \sim 0)
\sim R^{w_{J}}_{o} \).

The carbon-wax system is the same one as used earlier for noise measurements
in the range $p \ge p_c$\cite {nmb} but samples used in this work were in
the Joule range $p > p_J$. The fraction $p_J > p_c$ is such that the
resistance of a sample in response of an applied field increases if $p >
p_J$. That the increase in resistance in this range of $p$ is indeed due to
the Joule heating has been confirmed by a number of observations\cite {rem4}
like the sensitivity of measured values of resistances to the presence of a
cooling fan near samples. For this system, $p_c = 0.76\%$ by volume, $p_J <
3\%$ and $t=2.05 \pm 0.15$\cite {cbb}. C-W samples were all disk-shaped, 10
mm in diameter and 6 mm high. Connections to electrodes were such that
current flowed parallel to the thickness of the disc. The preparation and
characterisation of C-HDPE samples have been described in ref. \cite
{heaney}. $t$ has the value of $2.9 \pm 0.1$ in this system. These samples
had dimensions $10 \times 4 \times 1 $ mm$^3$. In this case, currents flowed
parallel to the longest side. All measurements were done at room temperature
at constant currents. For any constant current, sufficient time (more than
40 min) was allowed for the sample resistance to attain a steady value.
Furthur details of noise measurements can be found in ref. \onlinecite{nmb}.

Let $S_V$ denote the spectral power of voltage fluctuations at a constant
current. Fig. 1 shows plots of $S_V(0.5Hz)$ vs. $I$ for five C-W samples
with $p$ as indicated. At low currents (ohmic regime), $S_V$ varied as $I^2$
(solid straight lines). At high currents, $S_V$ varied faster than $I^2$ in
the similar manner as $V-I$ curves become nonlinear. Such concurrent
nonlinear behaviour of noise\cite {rem3} and resistance becomes more evident
in Fig. 2 where the normalised relative noise ${\cal S}$, defined by ${\cal
S} = S_V/ {V^2}$, (squares) and conductance $\sigma = 1/R$ (circles) are
shown as functions of the current for the two C-W samples with $p=$3.5
(closed symbols) and 10$\%$ (open symbols). The conductance rather than
resistance was used to avoid overlapping of data. The relative noise and the
resistance of a sample have similar qualitative behaviour in that both
remain close to their respective zero-bias values at small currents but
appear to deviate from those values as the latter is increased. The
currents, $I_s$ and $I_r$, at which onset of nonlinearity took place were
determined by two independent methods. In the first method, they were
determined by adopting the criteria that corresponded to the increase in
noise by 100$\%$ and the decrease in conductance by 5$\%$. These levels of
changes are indicated by the dotted lines in Fig. 2. The criterion of
100$\%$ change of noise level was adopted to increase the accuracy of the
determination of $I_s$ as the noise increased steeply by several orders of
magnitude while the conductance decreased only by an order of unity. As
$p~(R_o)$ increases (decreases), both $I_s$ and $I_r$ increase. Log-log
plots of both $I_r$ (circles) and $I_s$ (squares) vs. $R_o$ in the two
systems are shown in Fig. 3. Straight lines indicate power-law fits with the
slopes as indicated. In the second method, the currents for each sample was
scaled with respect to $I_s$ determined by trial and error such that the
curves ${\cal S}(I)/{\cal S}(0)$ vs. $I/I_s$ of all samples of each system
collapsed on a single curve\cite {rem4}. Similar procedures were followed
with resistance data for $I_r$ (see Fig. 3 of ref. \onlinecite {mbh}). This
method yielded $x_r = 0.53 \pm 0.02$ and $0.45 \pm 0.02$ and $x_s = 0.81 \pm
0.03$ and $0.68 \pm 0.04$ in C-W and C-HDPE respectively. Averages of $x_s$
from the two methods are shown in (\ref {eq:xs_exp}) and averages of $x_r$
were used in (\ref {eq:xs_the}). $w_J$ was obtained from the relative noise
in ohmic regimes, ${\cal S}_o$. Its value in C-HDPE was found to be $0.13
\pm 0.06$\cite {mbh}. But, the data of C-W had large scatter. As a
consequence, a power-law fit yielded $w_J = 0.11 \pm 0.21$\cite {rem4}.
These values in Joule samples, significantly smaller than those near $p_c$,
were used in (\ref {eq:xs_the}) to obtain finally the predicted values as
mentioned above.

\begin{figure}
\includegraphics{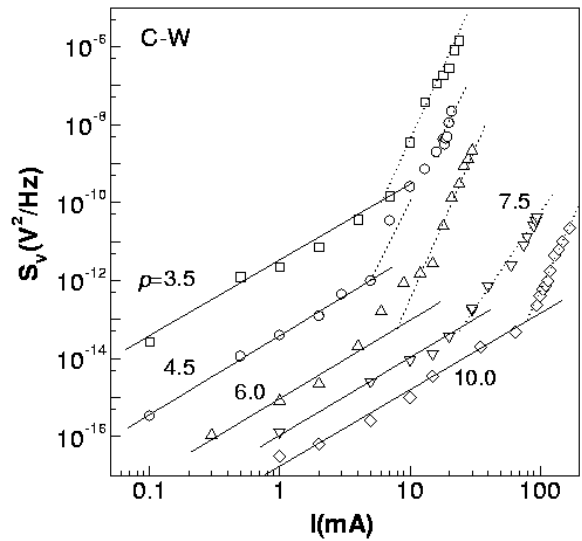}
\caption{Plots of noise power $S_V(0.5Hz)$ vs. dc current $I$ in five
carbon-wax samples with carbon percentage by volume ($p$) as marked. The
solid lines are fits to a power law with all slopes equal to 2 and the
dashed curves are only guide to the eye. For clarity, the data corresponding
to $p$ = 3.5 and 4.5$\%$ have been shifted upward by factors of 100 and 10
respectively.}
\end{figure}

The most significant feature of the nonlinear noise in the Joule regime is,
of course, the huge increase in magnitude compared to the small change in
sample resistance. This is furthur emphasised by plotting (curve b) the
relative noise in a Joule sample (C-W, $p=10\%$) against the bias-dependent
resistance $R$ in Fig. 4 which also shows the variation (curve a) of
relative noise in C-W with static disorder characterised by the linear
resistance $R \equiv R_o(p)$. Let's note that all measurements in this work
were carried out in the steady and reversible states $below$ breakdown\cite
{mbh} and hence, no change in the topology of percolating networks is
expected. It is then clear that small changes in resistances of the network
elements alone can not account for the large increase of the noise level.
Indeed we show below that temperature fluctuations which modulate
resistances are responsible for this phenomenon. The local rise in
temperature results in a change in the local resistance $r$ determined by
the temperature coefficient of resistivity, $\beta = (1/r)dr/dT$. The final
state under a constant current condition and a positive $\beta$ is
determined by the balance between the heat generated and the loss of heat to
the environment per unit time.

\begin{figure}
\includegraphics{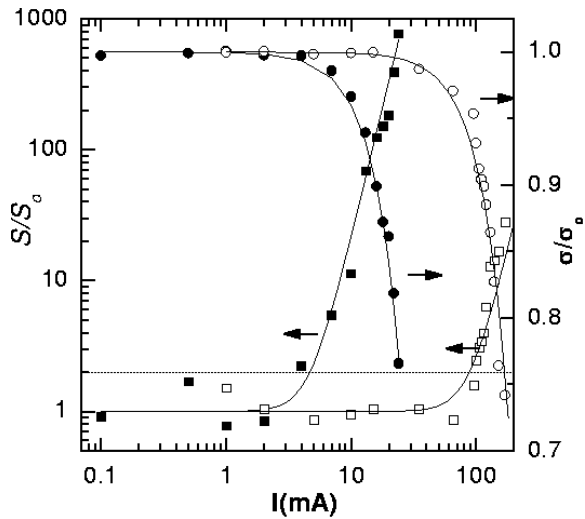}
\caption {Normalised relative noise power (squares) and conductance
(circles) are shown as functions of $I$ for two C-W samples with $p =3.5
{\%}$ (closed symbols) and $10.0{\%}$ (open symbols). The solid lines are
fits to the functionals ${\cal S} /{\cal S}(0) = 1 + a_s I^4$ and $\sigma /
\sigma_o = 1 - a_r I^2$ where $a_s(a_r) = 2.1\times 10^9 (400)$ and $1.7
\times 10^4 {\rm A}^{-4} (8.5 {\rm A}^{-2})$ for $p$=3.5 and $10{\%}$
respectively. The dashed lines indicate the criteria for determining the
current scales for onset of nonlinearity (see text).}
\end{figure}

The electrical response to Joule heating in percolating networks has been
treated in detail by several authors\cite {mbh,yagil,dubson}. In such a
network, the changed resistance due to Joule heating, in the first
approximation, is given by\cite {dubson}
\begin{equation} \label{eq:RI}
R=R_{o}+a \beta h R^{2}_{o} {\cal S}_o I^{2}
\end{equation}
Here \( a \) is a simple constant. \( h \) is the heat transfer coefficient,
defined as the ratio between temperature rise and power generated in a
conducting element. \( {\cal S}_o = {\cal S}(0) \) is the relative
noise power in the ohmic regime (Figs. 1 and 2). Each solid line through
conductance data in Fig. 2 is a fit to the expression $1 - a_r I^2$
according to Eq. \ref {eq:RI}. Here, $a_r = a \beta h R_o {\cal S}_o $. The
current scale $I_r$ for the onset of nonlinearity in resistance is given by
$I_r \sim { (a \beta h R_o {\cal S}_o) }^{-1/2} = {a_r}^{-1/2}$ \cite {mbh}.
It may be noted that the quantity $h$, as defined above, strives to provide,
in the spirit of mean-field approach, a reasonable description of heat
conduction in inhomogeneous media, which otherwise poses a formidable
problem. Ultimately, it governs the process of exchange of heat between the
sample and the environment\cite {yagil} and hence, is quite susceptible to
fluctuations.

\begin{figure}
\includegraphics{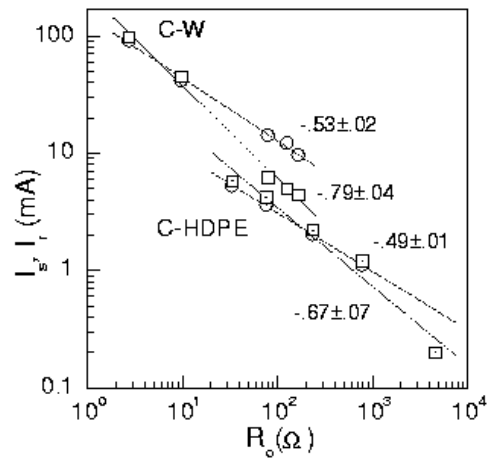}
\caption{Plots of nonlinearity current scales \( I_{s}\) (noise,
squares) and \( I_{r} \) (conductance, circles) vs. zero-field resistance \(
R_{o} \) for two systems. Data of C-HDPE samples have been shifted down by a
factor of 5. The solid lines are the power law fits to the data with the
exponents as indicated.}
\end{figure}

Eq. \ref {eq:RI} formed the basis for obtaining the $linear$ noise by the so
called third harmonic measurements\cite {yagil}. It will now be used to
calculate the noise in the $nonlinear$ regime. A small variation $\delta R$
is given by ${\delta R}/R = {(1+ a_r I^2 - 2 {a_r}^2 I^4 )} {\delta R_o}/R_o
+ a_r I^2 {(1 - a_r I^2)} {\delta h}/h $ after neglecting variation of
${\cal S}_o$. Using this and assuming that fluctuations in the ohmic
regime $R_o$ are uncorrelated with those in $h$ (or thermal
fluctuations), i.e. ${<\delta h \delta R_o>} = 0$, it could be easily
verified that ${\cal S}=<{{\delta R}^2}>/{R^2}$ is given by
\begin{equation} \label{eq:S1}
{\cal S} / {\cal S}_o \simeq 1 + 2 a_r I^2 + ({<\delta h^2>}/ {h^2
{\cal S}_o} -3) {a_r}^2 {I^4}
\end{equation}
after keeping terms up to the order of $I^4$. According to (\ref {eq:RI}),
$a_r I^2$ is equal to the fractional change of a sample resistance at
current $I$ which is of the order of 1 (see refs. \onlinecite {mbh} and
\onlinecite {nmb3} for its maximum possible values). Since the relative
noise, in contrast, increases in the same range of currents by several
orders of magnitude (see Fig. 2) we must have $ {<\delta h^2>}/ {h^2 \gg
{\cal S}_o}$ so that Eq. (\ref {eq:S1}) reduces to
\begin{equation}
\label{eq:SI} {\cal S} / {\cal S}_o \simeq 1 + ({<\delta h^2>}/{h^2 {\cal
S}_o}) {a_r}^2 {I^4}
\end{equation}
Fits to the noise data according to (\ref {eq:SI}) with $a_s = ({<\delta
h^2>} /{h^2 {\cal S}_o}) {a_r}^2 $ are shown in Fig. 2. The values of $a_s$
and $a_r$ given in Fig. 2 yield $ {<\delta h^2>}/ {h^2 {\cal S}_o} \sim
10^4~ {\rm to}~ 10^2$ in the range of 3.5-10$\%$ of $p$, much greater than
1. Goodness of fittings proves the earlier assertion that it is the
secondary (i.e. thermal) source of noise that is responsible for a huge
increase in noise level in the nonlinear regimes in Joule samples. The
primary source is, of course, the one responsible for ${\cal S}_o$. Let us
now consider the possible origin of such large magnitude of the thermal
noise.

\begin{figure}
\includegraphics{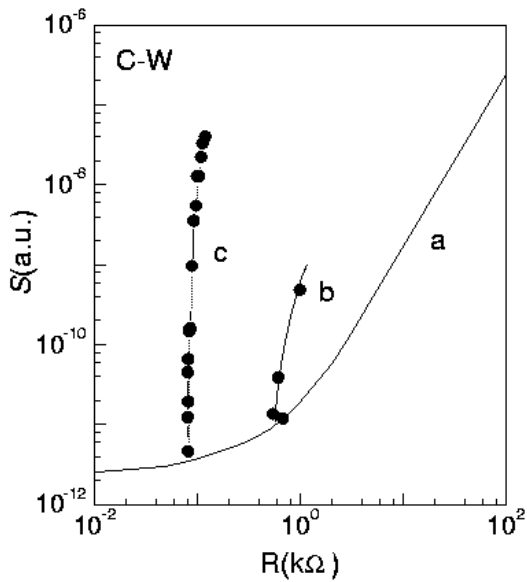}
\caption{Comparision of noise in nonohmic regime with that in ohmic regime
of a composite system. Relative noise is plotted against $R$ for two cases:
for curve (b), $R$ is the bias-dependent resistance of a joule sample ($p=10
\%$) and for curve (a), $R$ is the composition-dependent zero-bias
resistance. High resistance data of curve (a) are taken from ref.
\onlinecite {nmb}. }
\end{figure}

The inhomogeneous nature of percolating networks give rise to randomly
occuring 'hot spots' which are the regions carrying high current densities
and get heated most. Consequently, thermal gradients result in steady
heat flows away from those spots, thus setting up nonequilibrium steady
states. Now, a percolating system according to the 'Node-Link-Blob'
picture\cite {stauffer} could be viewed as a homogeneous system on the scale
of the correlation length $\xi(p)$. Links i.e. singly connected bonds in
this picture are the hot spots. With such multiple sources within the length
$\xi(p)$ and the fact that the fluctuation correlation length at equal times
in nonequilibrium states increases to macroscopic size\cite {schmitz} it
is assumed that a volume of $\xi^d$ will fluctuate thermally in a coherent
manner i.e. thermal $\xi_s$ will simply snap onto $\xi$. This means that
${<\delta h^2>} /h^2 \sim \xi^d$. Such coarsening of coherent length in
inhomogeneous systems explains the large increase in the thermal noise. It
is quite instructive to compare thermal fluctuation ${<\delta T^2>}/T^2$ in
the Joule regime with that in equilibrium. The latter at room temperature is
roughly inversely proportional to the number of particles in a sample i.e. 
$10^{-20}$ whereas the former is approximately equal to ${<\delta h^2>} /h^2
\sim 10^{-8}~ {\rm to}~ 10^{-10}$ with ${\cal S}_o \approx 10^{-12}$ at
0.5Hz. Note that in a homogeneous sample under a thermal gradient, $\xi_s$
can still be of microscopic size\cite {kogan}. From (\ref {eq:SI}), the
current scale for nonlinear noise is given by $I_s^{-1} \sim {({<\delta
h^2>} /{h^2 {\cal S}_o})}^ {1/4} {a_r}^{1/2}$ which, upon using $\xi \sim
R_o^{\nu /t}$ and $a_r = I_r^{-2}$, yields the relation (\ref {eq:xs_the}).
Note that the latter predicts $x_s - x_r$ to have theoretical values of 0.34
and 0.5 in 3D and 2D respectively ignoring $w_J$ and hence, the exponents to
be more divergent in two dimension.

In conclusion, we presented an experimental study involving two distinct
noise sources with different coherence lengths. In the present case, the
secondary thermal noise was induced by the Joule heating and became dominant
in the nonohmic regime due to amplification by coarsening of the coherence
length. This points to possible furthur use of noise study as a tool in
systems such as manganites\cite {tokura} which exhibit multiple phases
having very likely different fluctuation properties.

\begin{acknowledgments}
We are grateful to M. B. Heaney for the carbon-polyethylene samples. We
acknowledge the assistance of Arindam Chakrabarti in the preparation of C-W
samples.
\end{acknowledgments}


\end{document}